\documentclass[sn-mathphys-num]{sn-jnl}% Math and Physical Sciences Numbered Reference Style 

\usepackage{graphicx}
\usepackage{multirow}
\usepackage{amsmath,amssymb,amsfonts}
\usepackage{xcolor}
\usepackage{textcomp}
\usepackage{manyfoot}
\usepackage{booktabs}
\usepackage{algorithm}
\usepackage{algorithmicx}
\usepackage{algpseudocode}
\usepackage{listings}

\begin{document}

\title[A political radicalization framework based on Moral Foundations Theory]{A political radicalization framework based on Moral Foundations Theory}


\author*{\fnm{Ruben} \sur{Interian}}\email{ruben@ic.unicamp.br}

\affil{\orgdiv{Institute of Computing}, \orgname{University of Campinas (UNICAMP)}, \orgaddress{\city{Campinas}, \state{São Paulo}, \country{Brazil}}}

% \street{Av.~Albert Einstein~1251}, \postcode{13083-852}, 


\abstract{
Moral Foundations Theory proposes that individuals with conflicting political views base their behavior on different principles chosen from a small group of universal moral foundations. 
This study proposes using a set of widely accepted moral foundations (Fairness, Ingroup loyalty, Authority, and Purity) as proxies to determine the degree of radicalization of online communities. 
The fifth principle, Care, is generally surpassed by others, which are higher in the radicalized groups' moral hierarchy. 
Moreover, the presented data-driven methodological framework proposes an alternative way to measure whether a community complies with some moral principle or foundation: not evaluating its speech, but its behavior through interactions of its individuals, establishing a bridge between structural features of the interaction network and the intensity of communities' radicalization regarding the considered moral foundations. 
%from a social network or other online platform
%in order to derive structural properties that indicate the degree of correspondence between the current online community structure and \red{AAAAA} of each moral foundation. 
Two foundations may be assessed using the network's structural characteristics: Ingroup loyalty measured by group-level modularity, and Authority evaluated using group domination for detecting potential hierarchical substructures within the network. 
By analyzing the set of Pareto-optimal groups regarding a multidimensional moral relevance scale, the most radicalized communities are identified among those considered extreme in some of their attitudes or views. 
The application of the proposed framework is illustrated using real-world datasets. 
The radicalized communities' behavior exhibits increasing isolation, and its authorities and leaders show growing domination over their audience. 
There were also detected differences between users' behavior and speech, showing that individuals tend to share more `extreme' ingroup content than that they publish: extreme views get more likes on social media. 
}




\maketitle

\section{Introduction}

According to Cambridge Dictionary, morality is a set of standards for good or bad behavior; it is also the quality of being right, honest, or acceptable~\cite{defMorality2023}. 
Moral norms help us answer the question, ``How should we live?'' at the individual level. 

Politics, on the other hand, address the same question at the social level, establishing public policies and laws that reinforce some behaviors and discourage others. 
Despite the complexity and wealth of different political viewpoints, a single left-right (or liberal-conservative) continuum is generally used as a helpful approximation~\cite{2009Graham}. 
%, having predictive value for voting patterns and beliefs on various topics. 
Liberals generally have an optimistic view of human nature and the possibility of human perfection. 
They hold what Sowell~\cite{2007Sowell} refers to as an ``unconstrained vision'' according to which individuals should be given the greatest amount of freedom to follow their own paths in personal development. 
On the contrary, conservatives traditionally have a more pessimistic outlook on human nature, assuming they are naturally flawed, imperfect, or limited in their aspirations. 
As a result, they adhere to the so-called ``constrained vision'', according to which individuals must submit to the rules of institutions, authority, and tradition to coexist peacefully. 

The connection between moral and politics is reflected in the fact that individuals with different political views base their behavior on different sets of primary moral principles or foundations~\cite{2013Haidt}. 
%Moral foundations theory (MFT) is a social psychological theory that explains why morality exhibits a wide cultural variation while having several common characteristics and recurring themes. 
In a seminal article, Graham, Haidt, and Nosek~\cite{2009Graham} investigated how political liberals and conservatives construct their moral systems. 
They showed that in the U.S., liberals have essentially two psychological foundations (Care and Fairness), while political conservatives build their moral systems more evenly using five psychological foundations (the same ones as liberals plus binding foundations: Ingroup loyalty, Authority, and Purity). 

The existence of these five basic moral foundations (Care, Fairness, Ingroup loyalty, Authority, and Purity) was introduced in a widespread psychological framework called Moral Foundations Theory (MFT)~\cite{2004Haidt}. 
There is a vast literature applying MFT to moral psychology, bioethics, psychopathy, religion, and politics, and since its emergence, the central articles of MFT have been cited thousands of times. 

Studies showed that the positioning of individuals on the extremes of the left-right ideological spectrum is associated with their greater degree of radicalization~\cite{2021Santos}. 
%The more extreme an individual's opinion on the left-right ideological spectrum, the greater their degree of radicalization~\cite{2021Santos}. 
Radicalization is also defined as the process of changing the individual's beliefs and behaviors that involve a growing acceptance of the use of offline or online violence to achieve their ideological goals~\cite{2022Ellefsen,2008McCauley}. % political or ideological
%\red{However, there is no consensus on the violent nature of all radicalized communities, and radicalization can also result in nonviolent but dangerous actions (e.g., self-harming religious sects).} 
The increasing societal polarization is directly related to the radicalization process of each individual's views observed in various online communities. % ~\cite{2021Santos}

The growing polarization and radicalization of individuals and groups is an increasingly relevant social threat~\cite{2018Interian}. 
The authors of the most recent Global Risks Report~\cite{2023wef} published by the World Economic Forum (WEF) state that the erosion of social cohesion and polarization ranked as the most significant societal risk among those pointed out by WEF 2023. 
Moreover, societal polarization is listed as the fifth-most severe global risk in the short term. 

The goal of this article is to explore a novel characterization of radicalized online communities by positioning these groups over the multidimensional relevance scale of a set of primary moral foundations. 
Each moral principle or foundation corresponds to one dimension or axis of this scale, and the degree to which the individuals from some community 
%in the interaction network 
comply with this principle corresponds to its location on the axis. 

These specific research questions will be answered in this study: 
\begin{itemize}
    \item Is there a way to measure whether an online community complies with a moral principle or foundation by evaluating not only individuals' speech but also their behavior? % alternative [way], through interactions of individuals 
    \item How can online communities' radicalization be measured and compared, considering their different principles and moral foundations? % they are based on  
\end{itemize}

This study will explore whether two of the considered foundations -- Authority and Ingroup loyalty -- can be measured objectively by evaluating the group's behavior using the structural characteristics of the interaction network of individuals. 

Regarding the communities' radicalization measuring, the assumption being made in this study is that radicalized communities will be positioned at the extremes of the relevance scale for all or some of the considered moral foundations. 
The more extreme the positioning of the community on some specific moral principle, the more radical this community will be, and the more propensity it will have to act outside the legal norms to defend the principles it believes are absolute and unconditional. 

In a previous study, 
Bliuc et al.~\cite{2020Bliuc} stated that radicalized online communities are ideologically driven groups; therefore, their collective beliefs, values, and norms are crucial to understanding their actions. 
Hahn et al.~\cite{2019Hahn} applied the conceptual framework of moral foundations theory to discover the dominant motives and moral motivations of several specific terrorist groups. 
%They discovered that extremist right-wing ideologies were associated with a higher relevance of the binding moral motivations of Loyalty, Authority, and Purity. 
%, while extremist left-wing ideologies were associated with the individualizing motivations of care and fairness. 
Radicalization was also studied by identifying ``warning'' speech markers in online communities; however, this approach 
% does not consider the structure of interactions between individuals and groups, and 
is not very generalizable, given the difficulty of establishing ad hoc the appropriate and general set of warning radicalization markers that will work for any group~\cite{2019Grover}. 

This article is organized as follows. 
The next Section presents the methodology for defining the relevant set of primary moral foundations and establishing a data-driven relevance scale for them. 
It also proposes an approach to choosing the most radicalized communities using this multidimensional scale based on the content they produce and their structural characteristics. 
The results that illustrate the practical application of the proposed framework are presented in Section~\ref{secRES}. 
Conclusions are drawn in the last Section. 

\section{Methods}

This section describes a data-driven methodology that defines the positioning of communities on the relevance scales of a set of primary moral foundations. 
This methodology is based on analyzing the individuals' speech and their interaction networks in online platforms, where people generally are grouped in several communities of like-minded users. 
%The positioning of some community on the multidimensional moral foundations' scale is defined using the content people produce and the interaction network's structure. 

For each moral foundation, a scale will be established that indicates the degree to which the group complies with this principle. 
%expected behavior to each online community's real structure. 
Given a set of communities, an approach to choose the most radicalized ones using the multidimensional scale of several moral foundations is also described in this section. 

\subsection{Choosing primary moral foundations for measuring radicalization}

Moral Foundations Theory states that there are five widely accepted moral foundations~\cite{2013Haidt}. 
The first two foundations, Care and Fairness, are called the individualizing foundations because they emphasize the rights and welfare of individuals. 
The other three foundations, Ingroup loyalty, Authority, and Purity, are called the binding foundations because they emphasize group-binding loyalty, duty to the group, and self-correspondence to group ideals. 
Further studies sometimes included a sixth moral foundation: Liberty/oppression~\cite{2012Iyer}, although there is no consensus regarding its ``foundationhood''~\cite{MFT2023}. 
In this article, it will be assumed the existence of the five consensual moral foundations. 

However, since radicalization involves a growing acceptance of the use of violence~\cite{2022Ellefsen,2008McCauley}, Care foundation will not be considered in our framework. 
Human behavior is driven by moral hierarchies in which some motivations are deemed more important than others. 
Even though humans have compassion-driven psychological restraints against harming others, Bandura~\cite{1990Bandura} showed that radicalized groups harm others, seeing their actions as an effort to uphold moral values they deem most important, i.e., committing violence and rejecting concerns of Care in pursuit of ``broader aims''.
This means that the moral foundation of Care is surpassed by other principles that are higher in their moral hierarchy. 

%\red{\st{However, studies show that liberals generally have just two main psychological foundations (Care and Fairness), while conservatives build their moral systems more evenly using all five foundations. As a consequence, there are two foundations shared by people on both sides of the liberal-conservative (or left-right) continuum and three foundations that conservatives and liberals generally don't share. Moreover, the individualizing foundations of Harm and Fairness are not related to the collective behavior inside groups or communities.}} 
% \cite{2009Graham} \cite{2019Hahn}

%\red{Por outro lado, ``Fairness'' gerou revoluções!} 
%Purity was featured in only 7.64\%~\cite{2019Hahn} of groups' statements, mas da para identificar. 

%\red{The PPT-US contained information on 143 terrorist organizations. Of the 143 organizations in the database, 134 (93.71\%) were associated with at least one of MFT's motivations. Of those groups that were associated with at least one motivation, 42 were associated with two motivations, and 4 were associated with three motivations. Regardless of ideology or region of origin, loyalty was featured in most groups’ statements (90; 62.50\%), followed by fairness being reflected in 49 groups' statements (34.03\%), care in 20 groups' statements (13.89\%), authority in 14 groups' statements (9.72\%), and purity in 11 groups' statements (7.64\%). -- purity is the less important!!!}

Identifying the degree of relevance of the other four foundations (Fairness, Ingroup loyalty, Authority, and Purity) to some community allows for measuring how extreme or radical this community is. 
It provides indicators of incentives the individuals of the community have to act outside social boundaries, the established social rules that most people in society agree are reasonable or should be respected. 
Thus, political radicalization is analyzed in more detail, outside the widely used binary left-right simplification, introducing several dimensions often evaluated simultaneously or implicitly. 

Hereinafter, this study focuses on Fairness, Ingroup loyalty, Authority, and Purity as proxies to determine the degree of communities' radicalization. 
%The appropriate and measurable relevance scales for each of the considered foundations are proposed based on identifying the content and the structural features of the interaction network of individuals from online platforms. 

\subsection{Relevance of moral foundations using group speech}
\label{sec:speech}

One of the most traditional and direct ways to measure the relevance of each moral foundation to the individuals from some group or community is through the analysis of group speech. 
Different online communities use specific words to create ``frames'' that make attitudes or decisions seem morally good or bad~\cite{2014Lakoff}. 

Quantitative content analysis~\cite{2019Silverman} of user posts can be used as an objective approach to examine online communities' endorsement of specific moral principles. 
It involves creating categories of words and calculating the frequencies of words from each category in a corpus of text, as proposed in previous studies~\cite{2009Graham,2018Garten,2020ARAQUE}. 
In this case, each category corresponds to one moral foundation. 
While there is some subjectivity when choosing the set of words in each category, the method proved efficient in identifying differences in the attention given to each moral principle across the different groups~\cite{2009Graham,2018Garten}. 
% produced by the community

The Moral Foundations Dictionary (MFD), available at~\cite{MFT2023}, is a widely used categorization of hundreds of keywords into moral categories based on the MFT. 
There are also non-English versions of MFD. 
For example, MFD-BR is the Brazilian-Portuguese version of MFD created by Carvalho et al.~\cite{2020Carvalho}. 
%, having more than 700 categorized foundation-related words. 

For each online community and each motal foundation, the frequencies of words from the foundation-related category in the corpus of individuals' publications may be used as a primary indicator of the positioning of this community in the foundation's relevance scale. 

\subsection{Interaction networks}

The analysis of group speech is widely used to evaluate moral foundations' relevance to communities. 
An alternative way this study proposes to measure whether the community complies with some moral principle or foundation is to explore its behavior through its individuals' interactions. 

The behavior of individuals in a group is reflected in their interactions with other individuals from their own and other communities. 
Interaction networks are mathematical abstractions that represent the structure of social relations in some time period. 
The interaction network is modeled as a graph $G=(V, E)$, being $V$ a set of $n$ vertices (representing individuals), and $E$ a set of $m$ edges (representing interactions). %An edge $a=(x, y)$ indicates that the user $x$ interacted one or more times with the user $y$. 

There were identified two foundations that can be measured through the analysis of the structure of the interaction network: Ingroup loyalty and Authority. 
Next, it will be shown how can be evaluated the degree to which a community complies with these principles using the interaction network's structure, establishing a bridge between structural features of the interaction network of individuals and the intensity of communities' radicalization regarding the considered moral foundations. 

Note that different platforms have different characteristics that shape user interactions. However, each platform has one or more categories of interactions that represent endorsement. Retweets on Twitter mostly represent endorsement, as well as shares or likes on Facebook. Consequently, the same strategy as in this work may be applied to any social network or platform (e.g., YouTube) where users can interact through endorsements or recommendations. In all these platforms, it is possible to build a behavioral interaction network based on links that represent support or endorsement of one user to content published by another.

\subsection{Group-binding loyalty and isolation}

According to Haidt, 
% ``The Righteous Mind'' 
people always seek ways to create competing groups and cohesive coalitions: the Ingroup loyalty foundation is a part of our innate tendency to self-preserve and adapt~\cite{2013Haidt}. 
%Modern sports are much about expanding our Loyalty foundation triggers so that people can bind themselves together to pursue harmless trophies. 

In one of the four studies conducted by Graham et al.~\cite{2009Graham}, the participants rated several moral judgment statements on a 6-point scale (from strongly disagree to strongly agree), where each statement instantiated or violated some specific abstract principle. 
%The study showed that conservatives were more concerned than liberals about statements related to Ingroup loyalty. 
For example, the relevance of the Ingroup loyalty principle to participants was evaluated by statements like the following: 
\begin{quote}
``When it comes to close friendships and romantic relationships, it is okay for people to seek out only members of their own ethnic or religious group.''\\
``Loyalty to one’s group is more important than one’s individual concerns.''
\end{quote}

Group-binding loyalty reinforces the interactions with members of the same community and discourages interactions with other groups. 
In the limit, it leads to the isolation of individuals from people coming from any other group except their own. 
Therefore, a structural indicator for the Ingroup loyalty foundation should measure the degree of group cohesion and its isolation from different groups. 

Note that modularity~\cite{2004Newman} is a well-known network-level indicator of internal cohesion and isolation of communities. 
Given a division of some network's vertices into some collection of groups, modularity reflects the concentration of edges within groups compared with a random distribution of edges between all vertices regardless of the groups. 
More formally, if a graph $G(V, E)$ with $|V|=n$ vertices and $|E|=m$ edges, have its vertex set $V$ partitioned into $k$ disjoint groups $\{A_1, A_2, \ldots, A_k\}$, modularity $Q$ is defined as: 

\begin{equation}
Q = \frac{1}{2m} \sum_{u,v \in V}(a_{uv} - \frac{d(u) d(v)}{2m}) \cdot \delta_{g_u g_v}, 
\label{eqn:modularity}
\end{equation}
\noindent
where $d(v)$ is the degree of vertex $v \in V$; 
$g_v$ the index of $v$'s group; 
$a_{uv} = 1$ if there is an edge between vertices $u$ and $v$, $0$ otherwise; 
and $\delta_{ij}$ is the Kronecker delta. 

However, modularity is not designed to evaluate group-level cohesion or isolation. 
For example, if we evaluate the modularity of the subgraph induced by some group $A_i$ (i.e., a subgraph that contains all vertices in $A_i$ and all edges between vertices in $A_i$), then the modularity value $Q$ for this single group will be equal to zero~\cite{2016Newman}. 
A previous study showed a lack of consolidated measures for evaluating the cohesion and isolation of network nodes at the group level~\cite{2022InterianRev}. 

To evaluate group cohesion and its isolation from other groups in the network, an approach that measures the relative contribution of some specific group $A_i$ to the overall modularity of the network is developed. 
Since $\delta_{g_u g_v} = 1$ if and only if the vertices $u$ and $v$ are from the same group (and is zero otherwise), the equation~(\ref{eqn:modularity}) that defines modularity $Q$ can be rewritten as follows: 

\begin{equation}
Q = \frac{1}{2m} \sum_{i=1}^{k} \; \sum_{u,v \in A_i}(a_{uv} - \frac{d(u) d(v)}{2m})
\label{eqn:modularity_variant}
\end{equation}

Suppose that one specific group $A_i$ is chosen. 
Consider one addend in equation~(\ref{eqn:modularity_variant}) corresponding to the group $A_i$. 
% (the same reasoning applies to each group $A_i = A_1, A_2, \ldots, A_k$)
Note that this term $\sum_{u,v \in A_i}(a_{uv} - d(u) d(v) / 2m)$ considers only pairs of vertices $u$ and $v$ that both belong to $A_i$, and $\sum a_{uv}$ represents the actual number of in-group edges. 
Moreover, $\sum d(u) d(v) / 2m$ represents the expected number of $A_i$'s in-group edges after rewiring or randomizing the edges in the network while preserving the degree of every vertex (randomization known as the configuration model). 
If group $A_i$ gives no more within-community edges than would be expected by random chance, then the whole term $\sum_{u,v \in A_i}(a_{uv} - d(u) d(v) / 2m)$ will be nullified. On the other hand, if group $A_i$ gives significantly more within-community edges than would be expected by random chance, then the contribution of this group to network modularity will be large. 

Therefore, the contribution $Q_i$ of the group $A_i$ to overall network's modularity $Q$ can be measured in the following way: 

$$
Q_i = \frac{1}{2m} \sum_{u,v \in A_i}(a_{uv} - \frac{d(u) d(v)}{2m})
$$

Note that $\sum_{i=1}^{k} Q_i = Q$, i.e., the sum of the contributions of all groups produces the overall network modularity. 

However, a relative (not an absolute) contribution, of the group $A_i$ to the network's modularity is being sought. 
The $d$-modularity $d_i$ of $G$ respect to the group $A_i$ is defined as: 

$$
d_i = \frac{Q_i}{Q},
$$

Thus, $d_i$ reflects the relative contribution of some specific group $A_i$ to the overall modularity of the network. 

\begin{figure}[t]
\centering
\includegraphics[width=0.6\textwidth]{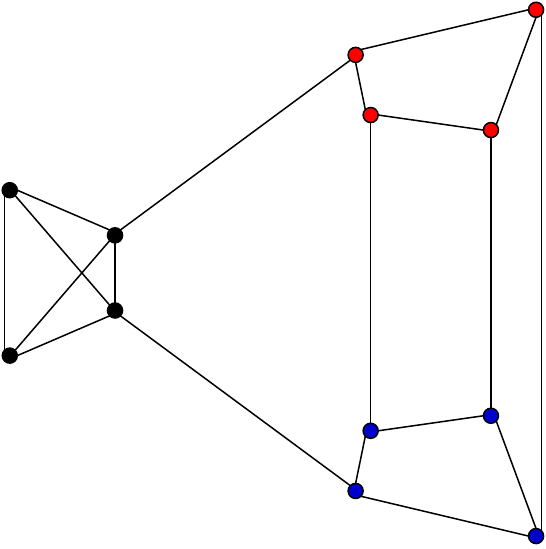}
\caption{Group-level modularity. Example of a graph with three groups, with the black group being much more internally cohesive and isolated.} 
%The contribution $Q_1$ of the black group is $0.180$ ($d_1=44.8\%$), while the contribution of red and blue groups is $Q_2=Q_3=0.111$ ($d_2=d_3=27.6\%$). 
% $Q = 0.402$
\label{fig:example_d_mod}
\end{figure}

Figure~\ref{fig:example_d_mod} shows an example of a graph with 12 vertices partitioned into three equally-sized groups, four vertices each. 
Note that each red and each blue vertex has exactly two in-group adjacent vertices and one adjacent vertex from a different group. 
Each black vertex, however, has exactly three in-group neighbors and, at most, one adjacent vertex from a different group: the black group is more internally cohesive and isolated. 
The overall modularity of the graph is $0.402$, and the group $A_1$ composed of black vertices has a relative contribution of $d_1 = 44.8\%$ to modularity ($Q_1 = 0.180$). 
The relative contribution of the other two groups, $A_2$ and $A_3$, with red and blue vertices, respectively, is quite smaller: $d_2=d_3=27.6\%$ ($Q_2=Q_3=0.111$). 

In summary, the $d$-modularity value $d_i$ defines the positioning of the community $A_i$ on the Ingroup loyalty foundation scale. 

\subsection{Authority and hierarchy}

Authority foundation 
%(and its antagonistic counterpart, subversion) 
is much about respecting (or not respecting) hierarchical relationships~\cite{2013Haidt}. 
These hierarchical relationships are generally bidirectional: up toward superiors and down toward subordinates. 
Authority should not be confused with raw power backed by the threat of force and the ability to inflict violence. 
% Of course, authorities often exploit their subordinates for their own benefit. However, 
Authorities in human societies are also about taking responsibility for maintaining order, justice, and representativeness in front of democratic institutions. 

In a previously cited study by Graham et al.~\cite{2009Graham}, participants evaluated the relevance of the Authority principle by rating moral judgment statements like the following ones: 
\begin{quote}
``If I were a soldier and disagreed with my commanding officer's orders, I would obey anyway because that is my duty.'' \\
``Respect for authority is something all children need to learn.''
\end{quote}

In summary, individuals guided by the authority principle are more sensitive to the sense of duty to the authorities and their group's leaders. 
In the limit, it means that all members of some group or community follow, 
%in their decisions, 
listen, or interact with a small set of authoritative voices and leaders. 
From a structural perspective, it means that the group shows a clear hierarchy among its members. 

\begin{figure}[t]
\centering
\includegraphics[width=0.75\textwidth]{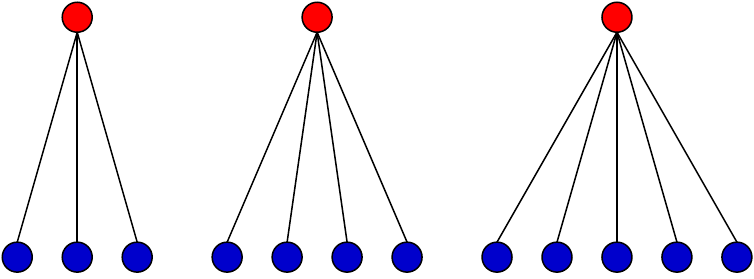}
\caption{Example of a fully hierarchical structure of relations.} 
\label{fig:example_hierar}
\end{figure}

Therefore, a structural measure for the Authority foundation should evaluate the degree to which some group shows a hierarchical structure. 
It may vary from the most disordered and chaotic (non-hierarchical) to a fully hierarchical structure of relations, as represented in Figure~\ref{fig:example_hierar}, which shows a network with 15 vertices, such as three red vertices (the ``authorities'') are able to reach all the remaining 12 blue vertices. 
As we shall see, identifying substructures like the one represented in Figure~\ref{fig:example_hierar} inside existing interaction networks is not always a computationally simple problem. 

Now, the goal is to identify a relatively small set of individuals (called authorities) capable of reaching (\textit{dominating}) the greatest number of individuals in the online interaction network. 
%These authoritative voices must be able to reach a large number of nodes representing the individuals inside some specific group. 
In short, the goal is to identify hierarchical substructures inside online interaction networks. 

For this purpose, the concept of domination is used. 
The mathematical study of domination comes from the 1960s. 
It was applied, for example, to identify coordinated communication in social networks by Campan et al.~\cite{2015Campan}. 
The reader is referred to Haynes et al.~\cite{1998Haynes} for a more detailed review of domination in networks. 

Since the \textit{degree} to which some group shows a hierarchical structure is to be measured, it is more appropriate in this study to use \textit{partial} domination. 
More formally, let $V = {1, \ldots, n}$ be the set of the $n$ individuals in the group (here, the interactions inside only one community are considered). 
The goal is to find the smallest subset $S^*$ of $V$ such that the number of individuals reached from $S^*$ 
%$|T \cup S^*|$ 
is at least $\rho \cdot n$, where $\rho$ is a parameter that reflects the minimum proportion of individuals to be reached. 
The problem can be studied considering, for example, $\rho=$ 100\%, $\rho=$ 75\%, or $\rho=$ 50\% of a given group. 

Note that when $\rho=$ 100\%, the above-mentioned problem is identical to the classical Dominating Set problem (DS)~\cite{1979Garey}. 
The smallest subset $S^* \subseteq V$ that reaches each vertex represents a group of authorities that dominates the whole group. 
The Dominating Set problem is known to be NP-hard, i.e., has no efficient algorithm for finding the optimum solution unless $P=NP$. 

However, in the more general case, when $\rho \leq$ 100\%, the problem requires a specific formulation called Partial Dominating Set problem (PDS). 
An instance of the PDS problem comprises a graph $G(V,E)$ and a parameter $\rho$. 
The goal is to find a dominating set that reaches at least ${\rho}|V| = {\rho}n$ vertices. 
The PDS is also NP-hard, since it contains the DS problem as a special case ($\rho = 1$). 

Since the problem is NP-hard, heuristic algorithms can be used to solve it on large interaction networks. 
Possibly the most acknowledged algorithm for the DS problem is a greedy constructive heuristic shown to be the best in terms of the approximation ratio for the DS problem unless $P = NP$ (for a more detailed analysis of the algorithm, see Parekh~\cite{1991Parekh}; for the approximation ratio proof, see Chlebík and Chlebíková~\cite{2004Chlebik}). 

At the beginning of the algorithm, $S^* = \emptyset$. 
Greedy adds a new vertex to $S^*$ in each iteration until $S^*$ forms a solution. 
In each iteration, it puts into $S^*$ the vertex (individual) connected with the maximum number of yet uncovered vertices and stops when all the vertices are covered. 
Note that the algorithm can be easily adapted for the PDS problem simply by stopping when the solution $S^*$ covers $ \rho |V| $ vertices. 

The lower the number of authoritative individuals in $S^*$, the higher the degree to which the group shows a hierarchical structure. 
This number, the cardinality of the set $S^*$, defines the positioning of the community on the Authority foundation scale. 
Note that groups with a non-hierarchical structure may need a very large number of individuals in $S^*$ to be dominated, a quantity of the magnitude of the size of the group itself. 

Finally, it is necessary to define how the value of $\rho$ should be chosen in practice. 
For example, Critical Mass Theory studies have found that rather large minority sizes (30\% or 40\% of the population) are often enough to trigger a collective behavior~\cite{2022Iacopini}. 
Nevertheless, in practice, the specific value chosen for $\rho$ generally affects the raw values, but does not significantly affect the \textit{relative positioning} of communities on the Authority foundation scale.

\subsection{Integrating the relevance scales}

Regardless of the type of relevance scales to be used (whether structural or speech-based), it is necessary to integrate them into one single decision-support framework that allows choosing the most radicalized communities. 

The problem of finding the most radicalized groups can be seen as a multicriteria optimization problem, where each moral foundation relevance is a criterion, and the goal is to find those groups that have the most extreme values across the multidimensional scale. 
In a situation where there are several entities (groups), each characterized by several variables (moral criteria), Multiple-Criteria Decision Analysis (MCDA) is a tool that can help find the solutions~\cite{2002Belton}. 

For example, let's consider the structural relevance scale composed of only two criteria: Ingroup loyalty and Authority. 
The criteria are evaluated in different ways. 
Ingroup loyalty is measured in a continuous $[0,1]$-based $d$-modularity scale, with greater (closer to one) values representing more isolation. 
Authority, on the other hand, is evaluated in a discrete integer scale representing the cardinalities of partial dominating sets, with smaller values representing better-defined hierarchical structures. 
Each criterion has an ``increasing radicalization'' direction: for Ingroup loyalty, more isolation (greater $d$-modularity) means more radicalization risk, and for Authority, smaller partial dominating sets represent an increasing radicalization risk. 

%\red{If we apply the speech-based approach, there are four criteria (Fairness, Ingroup loyalty, Authority, and Purity), the relevance scales are different (they are based on word frequencies), but the same reasoning is applied.} 

%\red{We have two options: (1) Maximize ALL three goals (corresponds to right-wing radicalized communities), or (2) Minimize ALL three goals (corresponds to left-wing radicalized communities). } 

Note that there may be groups that are no more radicalized than others according to all criteria. 
More formally, group $A$ is referred to as dominated by another group $B$ if and only if $B$ is as or more radicalized than $A$ with respect to all criteria. 
Pareto frontier, on the other hand, is a set of non-dominated groups. 
That is, there is no community in the Pareto frontier whose radicalization could be improved by another community without sacrificing at least one of the criteria. 
The Pareto frontier concept for MCDA is presented in detail, for example, by Luc~\cite{2008Luc}. 

The Pareto criterion is used when elements or solutions have several dimensions, and we try to find optimal elements following the assumption that all dimensions are significant and no one may be neglected. 
It offers an alternative to another common multicriteria approach of assigning (often) arbitrary or weakly justified weights to each criterion. 
In our structural relevance scale approach, the criteria can hardly be compared or combined using weights since Ingroup loyalty and Authority evaluation have very different natures. 

In the same way, the Pareto criterion may also be used in the speech-based approach. 
In this case, there are four dimensions, among which the communities are evaluated (Fairness, Ingroup loyalty, Authority, and Purity). 

When using the Pareto criterion adapted specifically for our radicalization decision analysis based on MFT, the goal is to choose the most radicalized communities among those considered extreme in some of their attitudes or views. 
The set of communities in the Pareto frontier are the candidates to be analyzed in more detail due to their radicalization risks.

\section{Results}
\label{secRES}

This section illustrates the practical application of the proposed framework using several real-world datasets. 
It is shown how, using structural relevance scales, radicalized communities can be detected by analyzing sets of non-Pareto-dominated groups. 
The framework also shows that sometimes group behavior aligns with group speech. 
In other cases, however, group behavior is disconnected from group speech. 

%\red{The goal is to illustrate how to use the proposed framework in practice.} 

\subsection{Datasets and communities} 
\label{secCSD}

The framework is evaluated using several datasets collected before, during, and after the 2022 Brazilian presidential election from the Twitter social network. 

The majority rule with two electoral rounds is used in the Brazilian presidential election. 
If no candidate gets the majority of votes in the first round, the two candidates with the most votes proceed to a second round, excluding all others. 
In 2022, the first round of voting was held on October 2. 
As no candidate for president received more than half of the valid votes, a runoff election was carried out on October 30. 

The data collection period was from September 19 to November 13, 2022, eight weeks overall. 
This period starts two weeks before the first round of voting and finishes two weeks after the second round. 

Each of the four created datasets covered two weeks of data, and contained all tweets that mention some of the election-related words in Portuguese (\textit{eleição OR eleições OR eleitoral OR eleitorais}). 
Each dataset is very large, containing millions of tweets and interactions, enabling the use of structural and speech-based approaches. 
Table~\ref{tab:datasets} shows the main features of the collected datasets, like the number of tweets and the number of tweet authors. 
The table also indicates which data covered which period: before, during, and after the election. 

\begin{table}[ht]
\caption{Datasets adopted for illustrating the use of the framework. Election days are marked in bold.}
      \begin{tabular*}{\textwidth}{@{\extracolsep\fill}lcccc}
        \hline
        Dataset & Dates & Period & Number of tweets & Number of users \\ \hline
        D1 & Sept. 19 -- \textbf{Oct. 2, 2022} & Before elections & 4,087,911 & 934,870 \\
        D2 & Oct. 3 -- Oct. 16, 2022 & During elections & 4,193,174  & 808,428 \\
        D3 & Oct. 17 -- \textbf{Oct. 30, 2022} & During elections & 8,131,875   & 1,126,346 \\ 
        D4 & Oct. 31 -- Nov. 13, 2022 & After elections & 6,264,584   & 1,025,486 \\ \hline
      \end{tabular*}
\label{tab:datasets}
\end{table}

The datasets contain a clearly radicalized right-leaning community that invaded the Supreme Federal Court, the National Congress building, and the Planalto Presidential Mansion in the Three Powers Plaza in Brasília on January 8, 2023. 
Several other communities were detected in the datasets. 

The community detection method by Blondel et al.~\cite{Blondel2008} was used for identifying in one run the communities for all four datasets, using the retweet network of interactions from September 17 to October 29, 2022, that is, approximately 75\% of the data. 
It was verified that starting from the election's second round day, October 30, 2022, the pattern of interactions was much a binary one.
Therefore, nearly 25\% of the data, starting from October 30, 2022, was excluded from community detection, so that binary voting decisions do not affect the communities formed during the whole period. 
The users who never interacted before the election's second round, i.e., those who only interacted after the electoral period, were not considered in the analysis of the fourth dataset. 

Several online communities were identified. 
Two specific groups (the left-leaning and right-leaning) can be easily positioned in the ideological spectrum by observing that many high-degree individuals in these groups corresponded to key political figures in the elections, such as the former president, Luiz Inácio Lula da Silva, and the current president at the time, Jair Bolsonaro. 
There were several other communities, the largest of which corresponded to mainstream media and press followers. 
We did not emphasize the left-right political identifications within the theoretical formulation of the framework. Still, many researchers and media widely use this left-right spectrum, and we used two specific communities of users that positioned themselves as adversarial groups traditionally placed on the extremes of this spectrum. 

Due to resolution limits in community detection in such large datasets~\cite{2007Fortunato}, only communities with at least 4000 individuals were considered, since communities smaller than $\sqrt{2L}$ users can result from an arbitrary merge of smaller structures, where $L$ is the total number of links in the network. 
Seven communities satisfied this restriction. 

\subsection{Radicalization by network structural features}

First, the radicalization of communities was evaluated considering only two structural criteria: Ingroup loyalty and Authority. 

\begin{figure}[h!]
\centering
\includegraphics[width=1\textwidth]{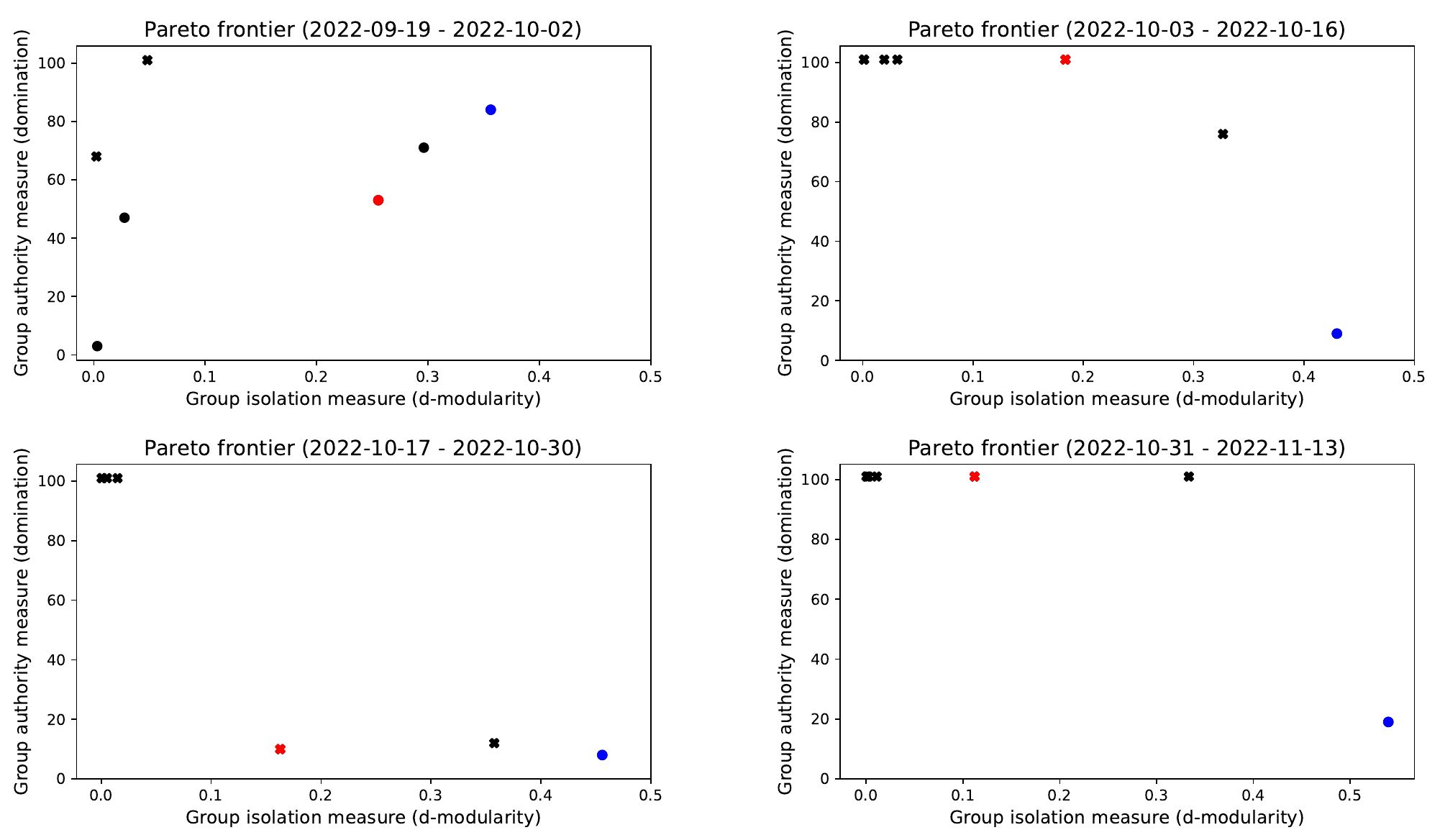}
\caption{Radicalization analysis using network structure. The Pareto frontier is calculated for two structural features. The color indicates the group leaning (red for left-leaning and blue for right-leaning), and the shape indicates the Pareto efficiency of the solution (circle \textbf{O} for Pareto-efficient and \textbf{X} mark for non-Pareto-efficient). } 
\label{fig:structural}
\end{figure}

Figure~\ref{fig:structural} shows the positioning of online communities in the two-dimensional scale for the four analyzed datasets. 
Each point corresponds to one community. 
The further to the right and down the point's positioning, the more radicalized that community is, indicating the ``increasing radicalization'' direction. 
The shape of the marker indicates Pareto-optimality (a circle \textbf{O} for Pareto-efficient and an \textbf{X} mark for non-Pareto-efficient), while the color of some groups indicates the group leaning (blue for right-leaning and red for left-leaning). 

The first dataset, containing data collected before the election's first round (October 2, 2022), did not show one but several non-dominated communities. 
However, the right-leaning group already appears in the Pareto frontier due to its relatively high Ingroup loyalty measure value. 
For the other three datasets, there is only one Pareto optimum. 
The right-leaning community dominates the rest of the groups, because its Ingroup isolation increases, while the dominating set sizes, which measure Authority strength, decrease simultaneously. 
The community's behavior exhibits increasing isolation, and its authorities and leaders show a growing domination over their audience. 
% with values of 0.36, 0.43, 0.46, and 0.55 for the four periods, respectively

The results show an increasing radicalization of the right-leaning community that invaded the three branches of government in Brasília on January 8, 2023. 
There are no other groups that exhibit this kind of behavior. 

\subsection{Radicalization by group speech measuring}

The relevance of moral principles to communities can also be measured using the more traditional speech-based method. 
However, also in this case, the multicriteria approach based on the Moral Foundations Theory can help indicate the most radicalized groups. 

Four dimensions are evaluated using the group speech of each community: Fairness, Ingroup loyalty, Authority, and Purity, as described in Section \ref{sec:speech}. 
The Brazilian-Portuguese version of MFD called MFD-BR created by Carvalho et al.~\cite{2020Carvalho} was used for the investigated datasets. 

For each of the seven communities, there are four values that represent the frequencies of the words related to each foundation in the corpus of texts published by this community. 
Parallel coordinates are used for visualizing these four-dimensional data. 
In this visualization, communities are represented as connected line segments. 
Each vertical line or axis represents one moral foundation measuring. 
Communities with similar measures tend to appear closer. 
Any group whose line appears on the top of some foundation's axis will belong to the Pareto frontier since no other group dominates it in this dimension. 

\begin{figure}[hb]
\centering
\includegraphics[width=1\textwidth]{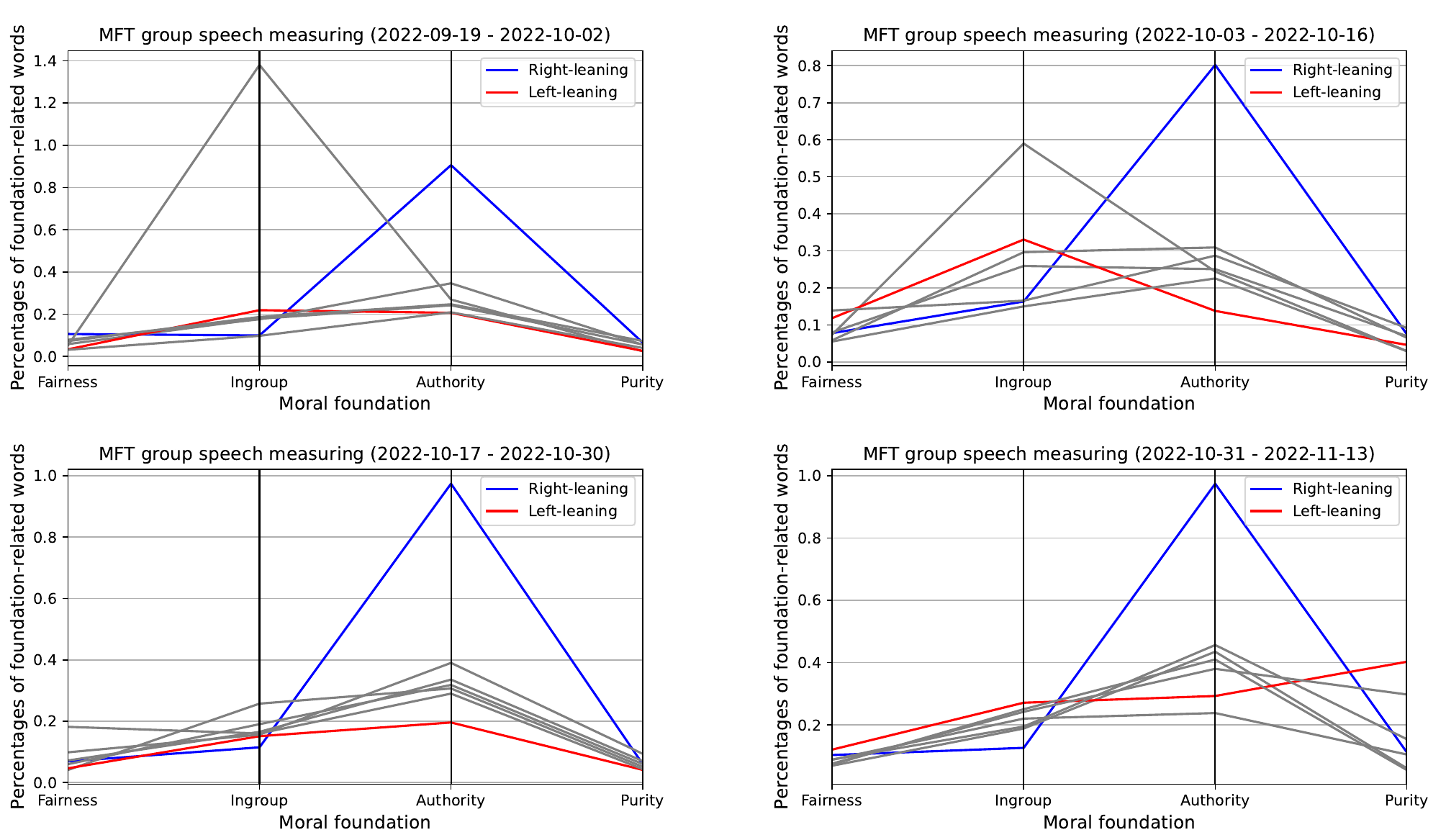}
\caption{Radicalization analysis using four group speech features. Communities are represented as connected line segments, and vertical lines represent moral foundations.} 
\label{fig:group_speech}
\end{figure}

Figure~\ref{fig:group_speech} shows the speech-based positioning of online communities for the four analyzed datasets. 
Looking at this visualization, it can be seen that the blue line greatly surpasses the rest in one axis, showing the extreme relevance of the Authority foundation for the right-leaning community when compared to the rest of the groups. 
This replicates the results obtained using the Authority structural feature in the previous section, showing that, in this case, group speech and behavior go hand in hand. 

However, speech-based Ingroup loyalty foundation measurement has a different behavior. 
The right-leaning community, more isolated and internally cohesive, does not beat the other groups, not showing a greater use of Ingroup loyalty-related words. 
The degree of isolation of the analyzed communities does not seem to be correlated with the use of these words, and group behavior is disconnected from speech-based Ingroup loyalty evaluation. 

The interpretation of the mismatch between behavior and speech regarding the Ingroup loyalty foundation may be firstly explained by specific circumstances in which the 2022 elections occurred in Brazil. Due to the existing polarization, the possible re-election of incumbent President Bolsonaro was interpreted by a significant part of the adversary electorate as a threat to the current political system. On the other side, for Bolsonaro’s followers, the possible election of former president Lula da Silva was also perceived as a nearly existential threat to their political group. 

In that context, the behavior we measure using interactions with content published by political groups strongly reflected this threat perceived by both communities: users strongly reinforced (endorsed) messages published by their political groups. However, on average, the content published by the same users was much more neutral, with fewer ingroup-related words; their speech had a larger diversity of content. 

In summary, our discovery implied that users promoted more strongly, on average, ingroup-oriented messages when compared to their average use of ingroup-oriented speech, which was much more neutral. These differences between behavior and speech show that users tend to share more “extreme” Ingroup content than that they publish: extreme views get more likes on social media \cite{2023Pandey}. 

In all four analyzed datasets, the other two moral foundations, Fairness and Purity, do not significantly differ in the frequency of foundation-related words across the groups. 
For Fairness, the frequencies of foundation-related words are similar in all cases. 
For Purity, they are very similar for all datasets, except for the last one, D4.

\section{Conclusions}

In this study, a political radicalization framework based on Moral Foundations Theory is presented. 
A novel characterization of radicalized online communities is explored by positioning these groups over the multidimensional relevance scale of a set of primary moral foundations. 

There are two ways to measure the degree to which individuals from some community comply with each moral foundation, i.e., the positioning of the community in the foundation's relevance scale. 
A more traditional method is based on evaluating group speech by measuring the appearance of foundation-related words in the content produced by individuals in the group. 

An alternative approach is evaluating group behavior in the network of interactions between individuals. 
Two foundations, Ingroup loyalty and Authority, may be measured using the interaction network's structural features. 
Using structural relevance scales, radicalized communities can be detected by analyzing sets of non-Pareto-dominated groups. 
The application of the proposed framework is illustrated using real-world datasets, with a radicalized right-leaning community that invaded the three branches of government in Brasília on January 8, 2023, being the only Pareto optimum for the last three (chronologically) of the four analyzed datasets. 

Therefore, the following answers were found to the research questions posed in this study: 
\begin{itemize}
    \item Is there a way to measure whether an online community complies with a moral principle or foundation by evaluating not only individuals' speech but also their behavior? \newline -- Yes, we founded two network features that, conceptually, reflect the degree to which individuals' interactions are consistent with their respective moral principles. 
    \item How can online communities' radicalization be measured and compared, considering their different principles and moral foundations? \newline -- Given a set of moral foundation's structural relevance scales, the set of communities in the Pareto frontier are the candidates to have the greater radicalization risk. We illustrated the use of the framework by showing that, unlike before the elections, during and after the 2022 Brazilian electoral process, the right-leaning radicalized community was the only Pareto optimum. 
\end{itemize}

Among the limitations of our study, the use of modularity-based community detection in interaction networks imposes resolution limits~\cite{2007Fortunato}, such that small groups (in our case, smaller than 4000 vertices) may go undetected. 
The focus of the study on large enough interaction groupings limits the potential of detecting smaller and more exclusive groups that are more easily radicalized and may lead to violence. We consider that in future studies, using community detection methods focused on smaller groups, such as hierarchical clustering, could facilitate discovering these smaller groups embedded into larger communities. 

In future works, we also intend to explore the idea of linking the moral foundations theory to levels of self-confidence and parenting styles that may potentially lead to a predictive model. 

In some cases, group behavior is in line with group speech. 
Regarding the Authority foundation, evaluated by detecting hierarchical structures within the network, there is a match between the presence of these substructures in communities and the use of Authority-related words. 

In other cases, group behavior is disconnected from group speech. 
For Ingroup loyalty foundation, the right-leaning community is more isolated and internally cohesive, despite its low use of Ingroup loyalty-related words when compared to other communities. 
The analyzed communities' isolation does not seem to be correlated with their use of words related to Ingroup loyalty foundation. 
These differences between behavior and speech show that users tend to share more `extreme' Ingroup content than that they publish: ``Extreme views get more likes on social media'' \cite{2023Pandey}. 

The behavior of radicalized communities detected through structural relevance scales indicates an increasing isolation and a growing domination of the authorities and leaders over their audience. 
The proposed framework can be used to identify those groups exhibiting risky behavior by analyzing the structural characteristics of social networks and other platforms of interacting users. 

%The other two moral foundations, Fairness and Purity, do not significantly differ in the frequency of foundation-related words across the four analyzed datasets. 
% -----------------------------------------------

\backmatter

\section*{Statements and Declarations}

\textbf{Competing interests} \ The author declares that he has no financial or non-financial competing interests. 

\bigskip
\noindent
\textbf{Funding} Ruben Interian was supported by research grant 2021/12456-5, São Paulo Research Foundation (FAPESP). 

\bigskip
\noindent
\textbf{Availability of data and materials} The dataset analyzed during the current study is available in the Mendeley Data repository~\cite{dataInterian2023}. More disaggregated data, including each Tweet's text, are available upon reasonable request from the authors. 

\bigskip
\noindent
\textbf{Acknowledgements} Research carried out using the computational resources of the Center for Mathematical Sciences Applied to Industry (CeMEAI) funded by FAPESP (grant 2013/07375-0).

%%=============================%%
%% If you are submitting to one of the Nature Portfolio journals, using the eJP submission system, please include the references within the manuscript file itself. You may do this by copying the reference list from your .bbl file, paste it into the main manuscript .tex file, and delete the associated \verb+\bibliography+ commands. 
%%=============================%%

\bibliography{sn-bibliography}

\end{document}